\documentclass[epj]{svjour}
\usepackage{amsmath}
\usepackage{amssymb}

\usepackage[final]{graphics}

\newcommand{\dv} { {\rm d} }

\newcommand{\yy} {\zeta}

\newcommand {\cs} {c_{\rm salt} }

\newcommand {\SRI} {\chi}

\begin{document}
\title{Polyelectrolyte multilayer formation: electrostatics and short-range interactions}
%
\author {Adi Shafir\thanks{\email{shafira@post.tau.ac.il}} \and
David Andelman\thanks{\email{andelman@post.tau.ac.il}}}
\institute{School of Physics and Astronomy,
    Raymond and Beverly Sackler Faculty of Exact Sciences,
    Tel Aviv University, Ramat Aviv, Tel Aviv 69978, Israel}
\bigskip
\date{04 December 2005}
\bigskip
\vskip 1truecm


\abstract{We investigate the phenomenon of multilayer formation via
layer-by-layer deposition of alternating charged polyelectrolytes.
Using mean-field theory, we find that a strong short-range
attraction between the two types of polymer chains is essential for
the formation of multilayers. For strong enough short-range
attraction, the adsorbed amount per layer increases (after an
initial decrease), and finally it stabilizes in the form of a
polyelectrolyte multilayer that can be repeated hundreds of times.
For weak short-range attraction between any two adjacent layers, the
adsorbed amount (per added layer) decays as the distance from the
surface increases, until the chains stop adsorbing altogether.  The
dependence of the threshold value of the short-range attraction as
function of the polymer charge fraction and salt concentration is
calculated.}

\maketitle

\section{Introduction}
\label{Intro} 

The study of polyelectrolyte (PE) chains interacting with charged
surfaces has generated a great deal of attention in recent years.
This interest arises, in part, because of the numerous biological
and industrial applications. The adsorption and depletion of
polyelectrolytes {on} charged surfaces have been extensively studied
using
analytical~\cite{wiegel,muthu,chatellier,varoqui1,varoqui2,joanny}
and numerical~\cite{itamar1,itamar2,itamar3,us1,us2} solutions of
the non-linear mean-field equations, scaling considerations
\cite{itamar1,itamar2,itamar3,us1,us2,manghi,borisov,netz,dobrynin,review,review1},
multi-Stern layers of discrete lattice models
\cite{vanderschee,fleer,vandesteeg,bohmer} and computer simulations
\cite{yamakov,messina1,muthu2,muthu3}.

In recent years, formation of polyelectrolyte multilayers {has} been
investigated
experimentally~\cite{decherbook,mohwald,mohwald1,mohwald2,mohwald3,mohwald4,decher,decher3,schmitt,loesche,caruso,caruso1,caruso2,regine,regine2,regine3,regine4,adamczyk,blomberg,cho,greene1,greene2}.
These multilayers are composed of alternating  positively and
negatively charged PEs and are constructed via a layer-by-layer
adsorption of polyelectrolyte chains, shown schematically in Fig.~1.
The first stage is to dip a charged surface  into a solution of
oppositely charged polyelectrolytes and salt. After the
polyelectrolyte chains adsorb on the charged surface, the surface is
taken out of the solution and washed in a clear water solution. The
washed surface and adsorbed layer are then placed in a solution of
another polyelectrolyte, of an opposite charge to the first PE chain
(see Fig.~1), and then washed again. This process can be repeated
for several hundred times and results in a PE multilayer
build-up~\cite{decher}. More recent studies~\cite{mohwald3,mohwald4}
have shown that these multilayers have interesting and potentially
useful applications both for planar and spherical geometries,
leading the way to creation of multilayered and hollow spherical
capsules.

\begin{figure}
 \resizebox{1.0\hsize}{!}{\includegraphics{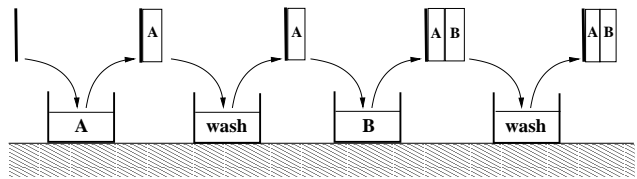}}
 \caption{A schematic illustration of the layer-by-layer PE deposition process.
    The charged surface is first dipped into a solution of oppositely charged
    PE chains (marked `A'). After the chains adsorb onto the
    charged surface, the surface is removed and dipped into a clear water
    solution (marked `wash'), removing any extra non-adsorbed chains. The surface is then
    dipped into another solution of PEs (marked `B'), carrying a charge opposite to the
    previously adsorbed PE chains, and washed
    again (`wash'). The process of dipping and washing: `A', wash, `B', wash, `A', wash,...
    can then be repeated for several hundred layers.}
\end{figure}

Theoretical models for  multilayer formation have also been
considered in recent
years~\cite{joanny,us2,castelnovo1,park,messina,qiang,patel}. In
previous studies~\cite{joanny,castelnovo1},  it was suggested that
the inversion of the surface charge by polyelectrolyte adsorption
occurs  under the following conditions: (i) sufficiently large salt
concentration; (ii) theta solvent conditions; and, (iii) weak
short-range (SR) interactions between the surface and the PE chains.
The charge inversion, together with complexation of polyanions and
polycations have been used to model the multilayer
formation~\cite{castelnovo1}.

In a previous publication~\cite{us2}, we reached different
conclusions by solving numerically the relevant mean-field equations
for the PE adsorption. We found that surface charge inversion will
not occur under a broad range of system parameters. More
specifically, it was shown that full charge inversion occurs only
for strong enough SR surface-PE interactions. Related results have
been reported in a separate study~\cite{qiang}, where the full
charge inversion was found to occur only for strongly hydrophobic
PE-backbone. Namely, PE chains in poor solvent conditions. The
strong hydrophobicity in the bulk creates an effective attraction to
the surface, replacing the bare surface-PE short-range attraction.


In the present work, we study the formation of alternating charged
PE multilayers as a function of polyelectrolyte charge, added salt
and SR interaction between the PE chains. We find that strong SR
(non-electrostatic in origin) interactions are necessary for the
formation of such multilayers, and that the adsorbed charge of the
alternating layers is not necessarily equal, or even close to, the
initial surface charge. Our analysis shows that the adsorbed amount
(per added layer) in the initially adsorbed layers always decreases.
If the SR interactions between the PE chains are too small to
attract another PE layer,  the multilayer formation will stop after
a small number of layers. However, if the SR attraction between the
alternating PE layers  is significant, the adsorbed amount (per
added layer) starts to increase back, and then saturates and forms a
stable multilayer stack. We also show how the multilayer formation
depends on the solution salinity and PE charge fraction.

In the next section, Sec.~\ref{mfeq}, we present the mean field
equations for multilayer formation and the~ numerical ~method used
to solve them. The numerical results for multilayer formation follow
in Sec.~\ref{res}, and their discussion in Sec.~\ref{discus}. We end
with conclusions and suggestions for future research in
Sec.~\ref{conc}.


\section{The Mean Field Equations}
\label{mfeq}


Consider an aqueous solution in contact with a bulk reservoir of
salt ions, and a dilute bulk concentration of long
polyelectrolyte chains. The solution is in contact with an
infinite and planar surface. The surface is oppositely charged
and attracts the PE chains. The mean-field equations for this
system were formulated in
Refs.~\cite{itamar1,itamar2,itamar3,us1}, and are repeated here.

\begin{equation}
  \frac{\dv^2 \yy}{\dv x^2}=\kappa^2\sinh\yy -4\pi l_B f \phi^2
 \label{PBmodSingle}
\end{equation}
\begin{equation}
  \frac{a^2}{6}\frac{\dv^2\phi}{\dv x^2}=v\phi^3+f\yy\phi+\omega^2\phi^5
 \label{EdmodSingle}
\end{equation}
$\phi^2$ is the local monomer concentration, $x$ the distance from
the charged surface, $\yy=e\psi/k_BT$ the renormalized
(dimensionless) electrostatic potential, $a$ the monomer size, and
$f$ the charge fraction of the PE monomers. The Debye-H\"{u}ckel
length $\kappa^{-1}\equiv \left(8\pi l_B \cs\right)^{-1/2}$ is the
screening length for electrostatic interactions in presence of
salt ions, and $l_B\equiv e^2/\varepsilon k_BT$ is the Bjerrum
length, which is approximately $7$\AA\, for water with
$\varepsilon=80$ and at room temperature.

Equation~(\ref{PBmodSingle}) is the Poisson-Boltzmann equation,
where the salt ions obey the Boltzmann distribution and together
with the monomer charges they act as charge sources for the
electrostatic potential. Equation~(\ref{EdmodSingle}) is the Edwards
equation for the monomer order parameter $\phi$, where the chains
are subject to an external potential composed of an electrostatic
potential and an excluded volume  interaction between the monomers.
Note that in Eq.~(\ref{EdmodSingle}) we included both the second
virial term modeled by  $v$ and the third virial one modeled by
$\omega^2$. In most previous studies of PE adsorption the third
virial term has been omitted, because of the dominance of the
electrostatic interactions and the second virial term. In the case
of multilayer formation, however, the third virial term becomes
significant, as is explained in the next section.  Finally, we note
that Eqs.~(\ref{PBmodSingle}) and (\ref{EdmodSingle}) are written
for vanishingly small bulk concentration of monomers, and under the
assumption that the ground state dominance approximation holds.

In order to model the build-up of multilayers, we note that the
experimental multilayer build-up is done via a layer-by-layer
adsorption of cationic and anionic PE chains, as is illustrated in
Fig.~1 and  explained in the introduction. During the adsorption of
each layer, the chains from previously adsorbed layers are believed
not to dissolve back into the {solution. Note} that the
layer-by-layer build-up is not a thermodynamically equilibrium
process. Any modeling of this phenomenon should take into account
these specific stages.

In our model, electrostatic and short-range (SR) interactions with
the PE chains of the previous layers are taken into account. The SR
interactions between the anionic and cationic PEs may have several
origins. The repulsive SR interactions include excluded volume
interactions, while an attractive SR interaction (beside the
electrostatic attraction) arises from {\it polyelectrolyte
complexation}. We do not offer in the present work a detailed
explanation for the complexation origin. We rather assume its
existence, which yields an effective SR attractive interaction, and
investigate under which conditions it will lead to multilayer
formation.  We assume that the electrostatic attraction and ion
pairing between the chains in the adsorbed multilayer and the 
adsorbing PE chains allow the adsorbing chains to penetrate 
the multilayer. This penetration slows down the PE chains
dissolution back into the solution, and creates an effective SR
attraction. This interaction has a non-equilibrium origin, but for
dense layers, it should last long enough to allow for multilayer 
formation.

The electrostatic interaction between the PE chains is taken into
account by adding the PE chain charges in the Poisson-Boltzmann
treatment, Eq.~(\ref{PBmodSingle}). The SR interactions between the
cationic and anionic PE chains are taken into account very simply by
adding a different interaction parameter $\SRI\ne v$ in
Eq.~(\ref{EdmodEq}). Hereafter, we assume that the charged fraction
$f$ of monomers for the negatively and positively charged PE chains
is the same. As the PE adsorption is done layer by layer, the
equations governing the adsorption of the $i$th layer are:

\begin{equation}
  \frac{\dv^2 \yy}{\dv x^2}=\kappa^2\sinh\yy -
    4\pi l_B f \left(z_i S_s+z_{i+1}S_o\right)
 \label{PBmodEq}
\end{equation}
\begin{equation}
  \frac{a^2}{6}\frac{\dv^2\phi_i}{\dv x^2}=
    v S_s\phi_i+  f z_i \yy\phi_i - \SRI S_o\phi_i
    +\omega^2S_s^2\phi_i
 \label{EdmodEq}
\end{equation}
where  $\zeta(x)$ is the electrostatic potential, and $\phi_i^2$ and
$f z_i$ denote the monomer concentration and monomer valency of the
$i$th layer, respectively.  The two above equations are solved
iteratively for the $i$th layer concentration $\phi_i$, while
assuming that monomer concentrations from all previously adsorbed
layers, $i-1, i-2, \dots, 1$ are fixed and known from previous
iterations. These SR interactions are contained in the two sums
appearing in the right hand side of Eqs.~(\ref{PBmodEq}) and
(\ref{EdmodEq}), $S_s$ and $S_o$. The sum
\begin{equation}
S_s\equiv\sum_{j=i,i-2\dots} \phi^2_j(x) \end{equation}
is the monomer concentration  at the point $x$, summed over all
similarly charged layers: $j=i, i-2, \dots$, which repel the
monomers of the $i$th layer. Similarly, the other sum:

\begin{equation}
S_o\equiv\sum_{j=i-1,i-3\dots}\phi_j^2(x)
\end{equation}
is summed over all oppositely charged layers having an attractive SR
interaction with the newly adsorbing PEs. It should be noted that
$S_s$ and $S_o$ are both functions of the layer number $i$, but the
subscript $i$ is omitted for simplicity. In the following, we
consider only monovalent PEs and set the odd layers as negatively
charged, $z_{2i+1}=-1$, whereas the even ones as positively charged,
$z_{2i}=1$.   We also note that for high monomer concentrations the
terms $S_0$ and $S_s$ are large, and thus higher orders of both the
excluded volume and the attractive interactions may be necessary.
However, in this simple model we restrict ourselves to the third
order only.

The solution of the pair of 2nd order differential equations,
Eqs.~(\ref{PBmodEq}) and (\ref{EdmodEq}), requires four boundary
conditions. Two of them are for the bulk where we choose
$\phi_i(x\rightarrow\infty)=0$, corresponding to a negligible amount
of PE in the bulk solution (dilute solution), and zero value for the
electrostatic potential, $\yy(x\rightarrow\infty)=0$. At the solid
surface, $x=0$, we use the electrostatic boundary condition
$\left.\dv\yy/\dv x\right|_{x=0}=-4\pi l_B\sigma$, where $\sigma$ is
the surface charge density.  For the first PE layer that adsorbs
directly onto the solid surface ($i=1$), we impose the Cahn-de
Gennes attractive  boundary condition~\cite{joanny,degennes}
$\left.\dv \ln(\phi)/\dv x\right|_{x=0}$ $=-d^{-1}$, where $d$ is a
characteristic length for the SR interactions between the surface
and the PE chains, and $d>0$ corresponds to an attractive surface.

For all subsequent layers, $i\ge 2$, we expect the PE chains to
partially penetrate into the previous layers because of the
complexation. In order to avoid the possibility of fully
interpenetrated layers, we introduce a hard wall for each layer at
an arbitrary location $x_i^*$.  Otherwise, in our case the cationic
and anionic polyelectrolytes would form a neutral complex, which
will not be able to form a stable PE multilayer. Because it is known
from experiments that the adsorbing PE chains of any specific layer
do not fully mix with the previous layers we introduce the concept
of the hard wall. Its justification would require further studies.
For the $i$th deposited layer, there is a (artificial) hard wall
$x_i^*$ inside the previous layers so that no monomers from the
$i$th layer can reach the region $x<x_i^*$. In order to simplify
notation, the layer index $i$ is dropped from the hard wall
notation, $x^*$.

The adsorption of every layer brings about a reversing of the
overall charge of the surface-PE-small ion complex. When the
adsorbed amount of ions and PE chains exactly balances the total
charge of the surface and previous adsorbed layers, the
electrostatic field perpendicular to the surface is exactly zero
(Gauss law). The hard wall of the adsorbing PE layer $x^*$ is taken
somewhat arbitrarily as the point where the electric field is zero.
As an example, the location of $x^*$ is depicted in Fig.~2.

Our choice of $x^*$ is motivated by our understanding the
complexation procedure. The driving forces for the adsorbing PEs to
penetrate the preceding layer are the electrostatic attraction
between oppositely charged PE chains and the ion pairing between
charged monomers. This electrostatic attraction is driven by an
attractive electric field for $x>x^*$. For $x<x^*$, the
electrostatic field repels the adsorbing PE, and no significant
complexation in that region is expected. Since we assume that the SR
attraction is a result of non-equilibrium complexation between
electrostatically attracted PE chains, we do not take $x^*$ to be
dependent on $\SRI$. It is important to note that such complexation
cannot occur between similarly charged PE chains, because the
repulsive interaction between the two chains as well as the excluded
volume repulsion would drive the chains to separate rather than
inter-penetrate.

\begin{figure}
 \resizebox{1.05\hsize}{!}{\includegraphics{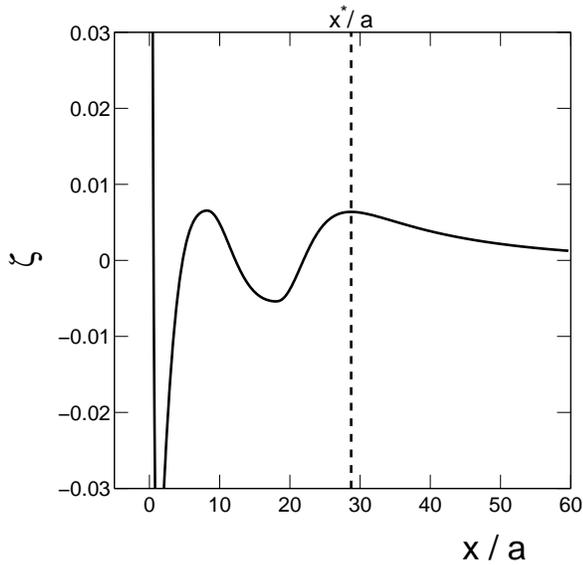}}
 \caption{The electrostatic potential
    $\yy$ (in dimensionless units) of  the first four alternating PE
    layers is presented as a function of the
    distance from the surface $x$. In each extremum point, the
    electrostatic field $\dv \yy/\dv x$
    changes sign, allowing the multilayer to attract an oppositely charged
    polyelectrolyte. The location of the farthest peak (marked by a dashed
    line) is taken as $x_5^*=x^*$  --- the hard wall condition for the
    interpenetration of the next (fifth) layer. This
    electrostatic potential was taken from the numerical profiles such as in
    Fig.~3. All parameter values are specified in Fig.~3. We note that
    the layers shown here are the initially adsorbed layers, and hence
    the strength of the potential oscillations is strongly affected
    by the existence of a charged wall. The amplitude of the electrostatic
    potential oscillations
    in the more distal region  increases towards the final layers,
    due to the increase in the adsorbed layer amount and charge.}
\end{figure}

The numerical procedure used to solve Eqs.~(\ref{PBmodSingle}) and
(\ref{EdmodSingle}), as applied to a single adsorbing PE layer, is
based on the relaxation method~\cite{NR}, and was presented in
detail in a previous publication~\cite{us1}. Here, we use the same
procedure for the layer-by-layer build-up. After obtaining the
solution for the first PE layer concentration, $\phi_1^2(x)$, and
its resulting potential, $\zeta(x)$, the layer monomer concentration
profile $\phi_1^2(x)$ is  frozen and added as a charge density
source to the right-hand-side of Eqs.~(\ref{PBmodEq},\ref{EdmodEq}).
These equations are now solved for the second layer using the hard
wall $x^*$ and the two virial coefficients, $v$ and $\SRI$. The
procedure is then repeated iteratively for all following layers in
order to obtain a multilayer stack.

\section{Results}
\label{res}

Our calculations show a strong dependence of the multilayer
formation on the value of the SR attraction coefficient $\SRI$ in
the case of a weakly good solvent, modeled via the $2^{\rm nd}$
virial coefficient $v=0.05a^3$ where $a$ is the monomer size. For
low amounts of added salt $\cs=0.1$M, the formation of multilayers
requires very large $\SRI/a^3\sim 3$ values, while for higher
amounts of salt $\cs=1$M the required $\SRI$ values drop to more
realistic values of $\SRI/a^3\sim 0.4-1$. For all salt
concentrations, low $\SRI$ values cause the adsorbed amount of
monomers in each layer to decay strongly with the layer number, so
that very few layers are formed. For high $\SRI$ values, the amount
of adsorbed monomers decreases for the initial layers and then
increases back. The adsorbed amount in each layer is found to reach
a stable value because of the third virial term. The threshold
$\SRI$ value is shown below to depend on the amount of salt in the
solution as well as the initial surface charge and the monomer
charged fraction.

The numerical solution of Eqs.~(\ref{PBmodEq}) and (\ref{EdmodEq})
yields the formation of multilayers, as presented in Fig.~3, under
the proper choice of parameters. As can be seen from the figure, the
multilayer can be divided into three spatial regions. In the
proximity region, containing the first few layers, the adsorbed
amount decreases substantially. In the intermediate region (layers
\,6-10), the monomer concentration increases rapidly to much higher
values. In the distal region, (under some conditions discussed
below) the adsorbed amount stabilizes, and the multilayer formation
continues. The adsorbed layers are shown to be very wide (of the
order of tens of nanometers) and highly concentrated. The
interpenetration between the layers looks to be quite significant.
The location  $x^*$, where the next layer begins to adsorb, is
shared by monomers from all four previous layers. This strong
interpenetration is the driving force of the multilayer formation,
since it allows for a strong interlayer SR attraction. Without it no
significant overcharging is achieved. The overall charge of each
adsorbed layer is much higher than the initial surface charge,
showing that there is no exact charge reversal in PE adsorption.

\begin{figure}
 \resizebox{1.05\hsize}{!}{\includegraphics{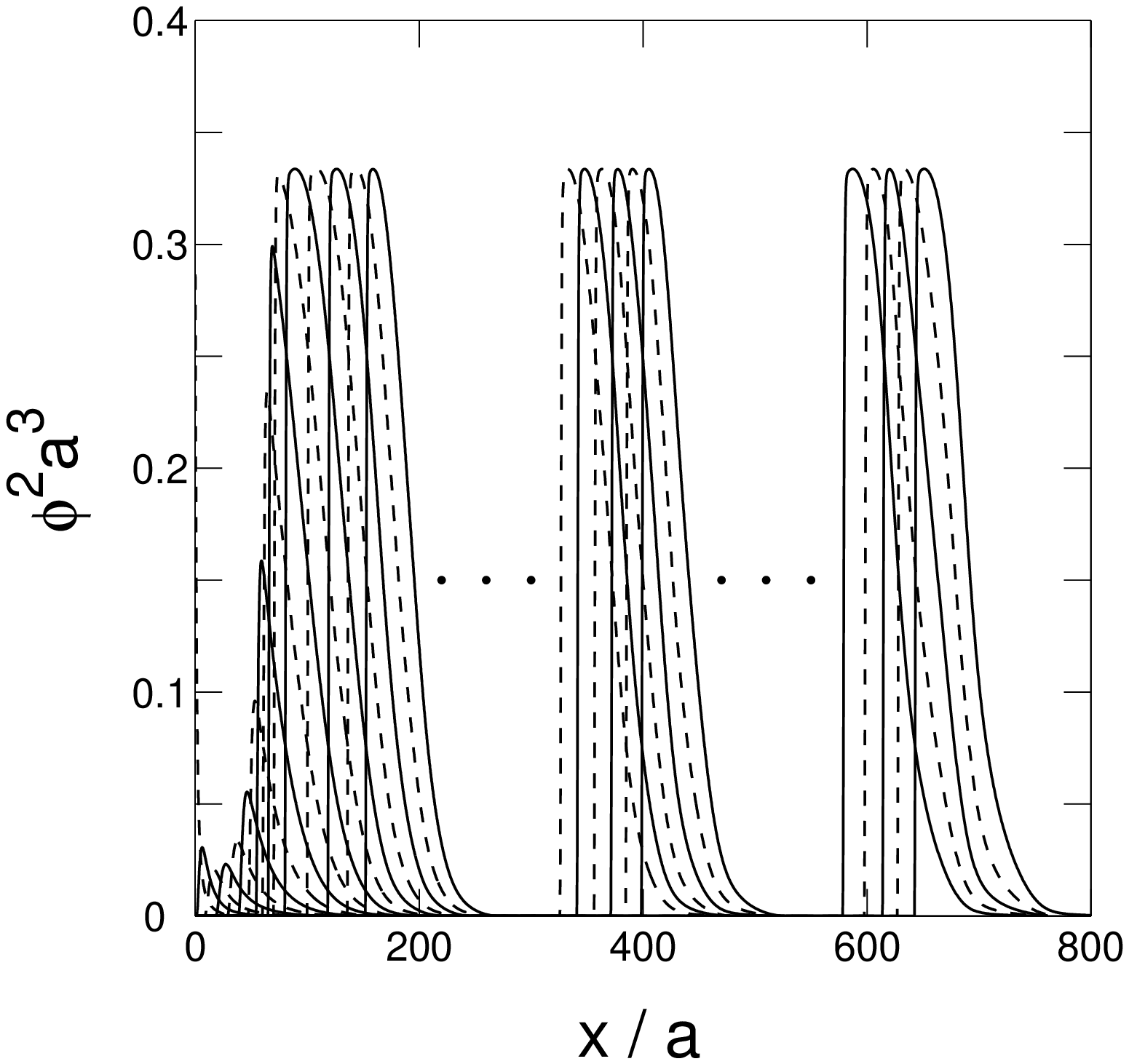}}
 \caption{The formation of a multilayer stack is shown for high $\SRI$ value, $\SRI=0.55a^3$,
  and the choice of parameters:
 $\cs=1.0$M, $a=10$\AA,
    $\sigma=2\cdot10^{-3}$\AA$^{-2}$, $d=50$\AA. For clarity purpose, we
    show only three groups of layers: layers  1-16 (left), 29-33 (middle) and 46-50 (right).
    For both polymers $v=0.05a^3$, $\omega^2=0.5a^6$, $f=0.5$.
    The aqueous solution has $\varepsilon=80$ and
    $T=300$K. The polycation profiles are marked by a solid line, while those
    of the
    polyanions by a dashed line. Three regions can be seen in the graph. Near the
    surface the multilayer concentration decays rapidly for layers 1-5
    (proximity region), and then increases rapidly in layers 6-10
    (intermediate region). The third
    region is where the multilayer concentration stabilizes. This
    stabilization occurs at a higher value than in the initially adsorbed
    layers. The layers in the distal region are highly
    interpenetrating, so that any layer interacts with about five other
    layers during its adsorption process. Note that the lowest
    monomer volume fraction in the proximity region is $0.025$, which
    is much lower than in the distal region. However, this layer
    is strongly complexated with the previous layers and should still be
    dense enough to survive the washing procedure.}
\end{figure}

The multilayer formation is characteristic of high $\SRI$ values. In
the opposite limit of low  $\SRI$ values, no stable multilayer stack
is formed because the complexation between the layers is not strong
enough. This case is shown in Fig.~4, which is obtained for similar
parameters as Fig.~3 except for a lower SR interaction coefficient
$\SRI$. The figure shows that the adsorbed amount in each
subsequential layer decays rapidly, until an additional layer cannot
be adsorbed, and the formation of a stable multilayer stack is not
possible.

\begin{figure}
 \resizebox{1.05\hsize}{!}{\includegraphics{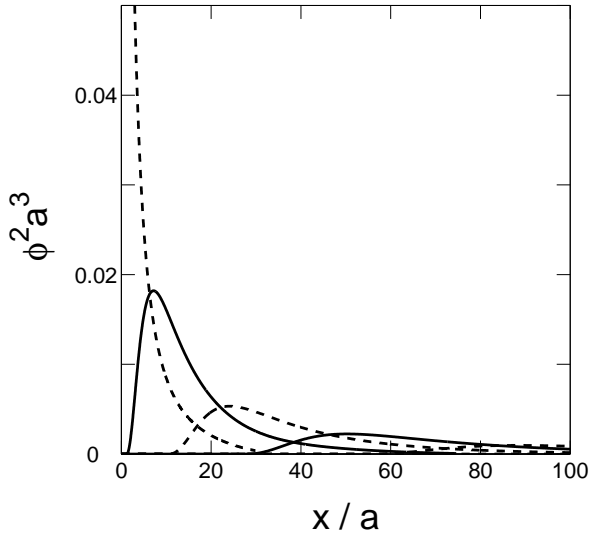}}
 \caption{The multilayer profile with lower $\SRI$ value,
 $\SRI=0.39a^3$,
 and all other parameters as in Fig.~3. For this low
    $\SRI$, the decrease in the non-electrostatic interaction between the
    PE chains causes the adsorbed layers to decay rapidly after four layers, and no
    stable multilayer stack is formed.}
\end{figure}

Within our model the formation of a stable multilayered stack
requires a third order virial term. Only when a strong enough
third-virial coefficient, of order $\omega^2\sim0.5a^6$, is added to
the SR interaction term, the multilayer concentration stabilizes at
high, but still physical, values. When the third virial coefficient
is too low, the adsorbed layer concentration does not saturate, and
rather reaches unrealistic high values. Our calculations also show
that an increase in the second virial coefficient is not enough to
stabilize the multilayers. It just drives up the threshold value of
$\SRI$. The spatial region where the adsorbed amount stabilizes is
the multilayer distal region, and it can be continued for as many as
80 layers (in our calculations) without any noticeable decay in the
adsorbed amount in each layer.

We end this section by showing three further results. In Fig.~5 we
show the overall thickness of the  adsorbed layers from Fig.~3 as a
function of layer number. In the mean-field model, the thickness of
the adsorbed layer is taken  as the position of the last monomer
concentration peak, which is a lower estimate  for the layer width.
The adsorbed layer width increases weakly for the first few layers
(proximity and intermediate regions), and then increases almost
linearly with the layer number (distal region) for the entire 50
layers. The linear increase shows that the multilayers are indeed
stable and reaches very high layer numbers.

\begin{figure}
 \resizebox{1.05\hsize}{!}{\includegraphics{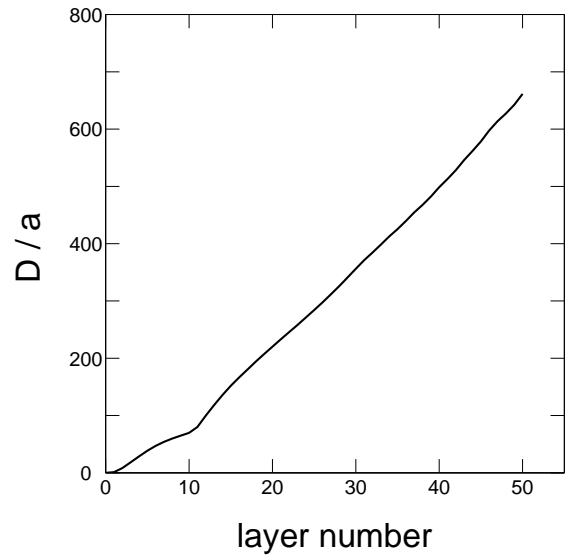}}
 \caption{ The total width $D$ of the adsorbed multilayer in Fig.~3
    is plotted as a function of the layer number. The (incremental) layer width is
    extracted from the peak position in the monomer concentration. The total width
    $D$ is seen to increase mildly for the initially adsorbed layers
    (proximity and intermediate regions), and then
    increase almost linearly, corresponding to the stable multilayer
    formation, with a constant thickness per each adsorbed layer.}
\end{figure}

In Fig.~6 we present the threshold strength of $\SRI$ that is needed
to form multilayers as a function of the added salt amount. As can
be seen from this graph, an increase in the amount of added salt
causes the necessary $\SRI$ to decrease strongly for low salt
concentrations. For higher salt concentrations the $\SRI$ value is
almost constant. The dependence of $\SRI$ on $\cs$,
$\SRI\sim\cs^\alpha$ fits roughly a power law with $\alpha\simeq
-0.8$, as can be seen in the inset of Fig.~6. However, this
empirical scaling is valid only for small range of salt
concentrations.

\begin{figure}
 \resizebox{1.05\hsize}{!}{\includegraphics{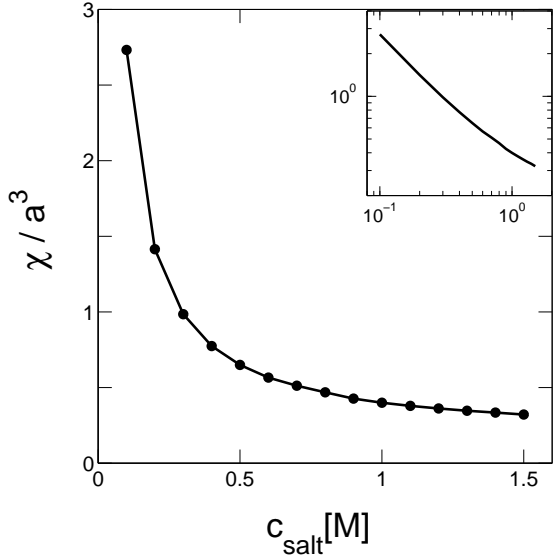}}
 \caption{The threshold value of $\SRI$  needed to create stable multilayers is plotted as a
    function of $\cs$. For low salt we get  high values of $\SRI\sim 3a^3$,
    while for high salt the value of $\SRI$ has lower values,
    around $0.5a^3$. All other parameters are
    as in Fig.~3, except  $d=10$\AA. The inset shows the same dependence on a log-log plot.
    For low $\cs$ values, the slope of the line
    can be fit to $\SRI\sim\cs^{-0.8}$, but for higher salt concentrations
    the exponent becomes lower. Since these changes occur over a
    single decade in $\cs$ values, there does not appear to be a good scaling
    law for $\SRI$ as function of $\cs$.}
\end{figure}

The dependence of the threshold $\SRI$ value on the mono\-mer
charged fraction $f$ is presented in Fig.~7.
 The $\SRI$ threshold increases with the
increase of $f$. Here, too, the numerical results do not
imply any simple scaling relation between $f$ and $\SRI$ (see
inset of Fig.~7).

\begin{figure}
 \resizebox{1.05\hsize}{!}{\includegraphics{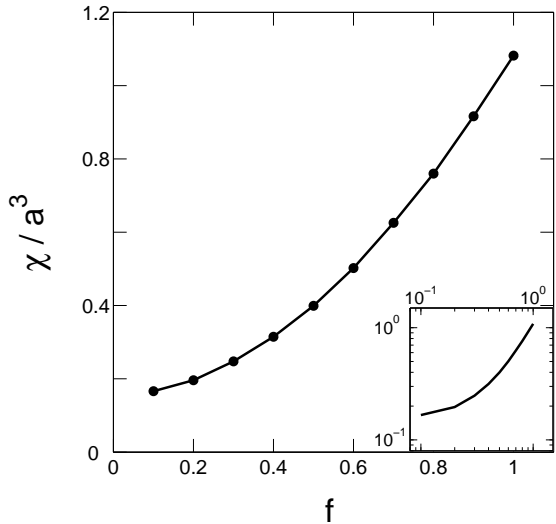}}
 \caption{The threshold value of $\SRI$ needed to create stable multilayers
 is plotted as a function
    of $f$. For low $f$ values, the monomer-monomer repulsion is low, and
    only a weak short-range attraction between the polymers is needed.
    For higher $f$ values the  threshold value of
    $\SRI$ increases. All other parameters are as in Fig.~3, except
    $d=10$\AA. The inset shows the same dependence on a log-log
    plot.
    No clear scaling law can be found for the dependence of
    the threshold of $\SRI$ on $f$.}
\end{figure}



\section{Discussion}
\label{discus}

The spatial behavior in Figs.~3 and 4 can be explained by the
following argument. In the proximity region, close to the surface,
the adsorbed layers have high monomer concentration and small width.
This small width does not allow for significant complexation between
the adjacent layers. Therefore, it causes the SR attraction between
the already adsorbed layers and the adsorbing PE chains of the
current layer to be low. This, in turn, causes a decrease in the
concentration of adsorbing monomers, accompanied by an increase in
the layer width.

The behavior of farther layers depends crucially on the strength of
the short-range attraction, $\SRI$. For low $\SRI$ values, the
monomer concentration in the farther layers continues to decay, and
the conditions are insufficient for multilayer formation. This
situation is shown in Fig.~4. In the opposite case of large enough
$\SRI$ values (depicted in Fig.~3), the increase of the layer width
causes the adsorbing polymers to interact with more than one
adsorbed layer. The complexation between adjacent layers becomes
stronger, allowing the monomer concentration in the adsorbing layer
to increase beyond that of previous layers (see intermediate region
of growth, layers 6-10 in Fig.~3). When the monomer concentration in
the adsorbing layer becomes high enough (layers 10 and above in
Fig.~3), the SR attraction between the different PE chains is
balanced by the third virial term of the excluded volume, and the
adsorbed amount in each additional layer stabilizes. This stable
multilayer is characteristic of the distal region, and persists to
dozens and even hundreds of layers without any noticeable decay.
We note that the specific built-up of the first dozen layers is a
direct consequence of our simple model of attractive interactions. A
more elaborate model may be needed for quantitative comparison with
experimental findings.

We now turn to the $\SRI$ threshold value needed to obtain a stable
multilayer formation. The threshold comes about because of the
competition between the SR and electrostatic interactions. The
decrease of the $\SRI$ threshold  with salt, as seen in Fig.~6, can
be explained in the following way.  Added salt screens the
electrostatic interactions and results in an increase in the PE
adsorbed amount and layer width~\cite{us2}. The thicker layers have
a larger contribution to the attractive SR term in
Eq.~\ref{EdmodEq}, and lead to a lower $\SRI$ threshold.

The dependence of the threshold $\SRI$ on $f$ (Fig.~7) is explained
qualitatively as follows. An increase in $f$ causes an increase in
the monomer-monomer repulsion between the adsorbing PE chains, and
an increase in the electrostatic attraction of the adsorbing PE
chains to the already adsorbed oppositely charged layer. The
increase in the threshold $\SRI$ with $f$ shows that the main effect
of increasing $f$ is to decrease the adsorption, meaning that the
main driving force of multilayer formation is not the charge
reversal caused by the adsorbing polymer layers, but rather the SR
interaction. It is important to note that in
experiment~\cite{regine2} a threshold $f$ value for the multilayer
formation was found. Above this threshold, the multilayer
concentration decreases with $f$, which is in agreement with our
findings. A threshold in the $f$ value is not found in our
calculations, mainly because our SR interactions are externally
imposed and do not depend on the value of $f$. However, the
threshold can be understood qualitatively as the value of $f$ for
which the polymer chains begin to interpenetrate, giving rise to the
SR attraction between them. We believe that further studies are
needed to better understand the origin of this threshold value.

Our simple model is subject to several limitations. First, we use
the mean-field theory and the ground-state dominance approximation,
valid for long PE chains.  The adsorption of short polyelectrolyte
chains requires other treatments such as molecular dynamics, Monte
Carlo simulations or lattice models. Second, during the adsorption
of each layer, we assume that all preceding layers are frozen,
meaning that they do not dissolve back into the solution or change
their spatial conformation. This assumption stems from the fact that
within mean-field theory there is no way to distinguish between an
adsorbed chain and a chain that is merely `stuck' at the surface
vicinity. This deficiency of mean-field does not allow us to give an
accurate model for the washing procedure. Experimentally, we expect
the polymer concentration to decrease, especially during the washing
step, as was modeled by other techniques~\cite{patel}. However, due
to long relaxation times we do not think that the washing will
affect drastically the structure of the already adsorbed multilayer
stack. Finally, we use a very simple model for the SR attraction,
and do not offer any explanation for its dependence on the PE
parameters. Despite these limitations, we believe that our model
gives an insight for the multilayer formation problem.

Quite recently, we have become aware of a alternative mechanism for
multilayer formation suggested by Q. Wang \cite{qiang2}. This
involves multilayer formation for poor solvent condition (balanced
by electrostatics). The multilayer formation can be achieved even if
the polyanion and polycation chains repel each other at short
distances, because the poor solvent condition induces stable
multilayer built-up. The solvent condition in Wang's model plays a
similar role as the attractive interaction between the cationic and
anionic chains in our model.

\section{Conclusions}
\label{conc}

We present a model aiming to explain PE multilayer formation for
marginally good solvents. The model is based on strong enough
short-range interactions between the polyanion and polycation. This
strong short-range interaction is shown to be indispensable for our
modeling of  such stable multilayers. We show that the multilayers
form easily in high ionic strength conditions, and that their
formation does not rely exclusively on the electrostatic attraction
to the previously adsorbed layers. We also calculate what is the threshold
strength of the short-range interactions needed for the formation of
multilayers as a function of the salinity as well as monomer charge
fraction.

In our model the multilayers are quite thick and interpenetrating,
while in the experiments the layers are thinner. However, we believe
that this simple model gives good insight on the problem of
multilayer formation, and can serve as a starting point for more
refined models. A possible extension will be to use a more specific
model for the short-range interactions between the PE chains, which
may give a better explanation to the experimentally observed
multilayer formation.

\begin{acknowledgement}
The authors would like to thank Alexander Grosberg, Jacob Klein, Henri 
Orland and Qiang (David) Wang for benefiting discussions. 
Support from the Israel Science Foundation 
(ISF) under grant no.
160/05 and the US-Israel Binational Foundation (BSF) under grant no.
287/02 is gratefully acknowledged.
\end{acknowledgement}


\end{document}